\documentclass[letterpaper,twocolumn,10pt]{article}
\usepackage{usenix}

\usepackage{tikz}
\usepackage{amsmath}
\usepackage{graphics}

\usepackage{cite}               
\usepackage[]{hyperref}         
\hypersetup{                    
  colorlinks,
  linkcolor={black},
  citecolor={black},
  urlcolor={black}
}
\usepackage{xurl}                
\usepackage{xcolor}             

\graphicspath{ {graphics/} }

\usepackage[font=small]{caption}
\usepackage{subcaption}

\usetikzlibrary{arrows,positioning}

\usepackage{xcolor}
\usepackage{comment}

\begin{document}

\date{}

\title{\Large \bf Millimeter-Wave Automotive Radar Spoofing}

\author{
{\rm Mihai Ordean, Flavio D. Garcia}\\
\{m.ordean, f.garcia\}@cs.bham.ac.uk
\\
University of Birmingham
} 

\maketitle

\begin{abstract}
Millimeter-wave radar systems are one of the core components of the safety-critical Advanced Driver Assistant System (ADAS) of a modern vehicle. Due to their ability to operate efficiently despite bad weather conditions and poor visibility, they are often the only reliable sensor a car has to detect and evaluate potential dangers in the surrounding environment.

In this paper, we propose several attacks against automotive radars for the purposes of assessing their reliability in real-world scenarios. Using COTS hardware, we are able to successfully interfere with automotive-grade FMCW radars operating in the commonly used 77GHz frequency band, deployed in real-world, truly wireless environments. Our strongest type of interference is able to trick the victim into detecting virtual (moving) objects. We also extend this attack with a novel method that leverages noise to remove real-world objects, thus complementing the aforementioned object spoofing attack. We evaluate the viability of our attacks in two ways. First, we establish a baseline by implementing and evaluating an unrealistically powerful adversary which requires synchronization to the victim in a limited setup that uses wire-based chirp synchronization. Later, we implement, for the first time, a truly wireless attack that evaluates a weaker but realistic adversary which is \textit{non-synchronized} and does not require any adjustment feedback from the victim. Finally, we provide theoretical fundamentals for our findings, and discuss the efficiency of potential countermeasures against the proposed attacks. We plan to release our software as open-source.
\end{abstract}

\section{Introduction}

Radar systems have been part of military and large commercial applications for decades, however, in recent years they have also become integral components of modern cars, e.g., in adaptive cruise control systems. Together with high-resolution cameras and LiDARs, millimeter-wave (mmWave) radars are the sensors through which modern vehicles perceive the surrounding environment and are an integral safety-related component of Advanced Driver Assistant Systems (ADAS). As part of ADAS, mmWave radars assist with object identification and tracking in situations where the other sensors are unable to perform optimally, such as in difficult light conditions or in the absence of direct line-of-sight. As such, for safety reasons, it is important for radars to produce accurate measurements that represent true depictions of the environment.

In recent years, the safety and reliability of sensors has come under scrutiny, mostly in the context of autonomous vehicles. LiDAR and camera systems have been thoroughly studied, and numerous successful object spoofing attacks have been demonstrated \cite{shin2017illusion,petit2015remote,sun2020towards,lidar_attack1,lidar_attack2,access_20,fung2017sensor}. However, attacks against radar systems, and especially the \textit{frequency-modulated continuous wave} (FMCW) radars used in cars, face fundamentally different challenges to LiDARs, and, have only been studied in artificial, either simulated or wired, environments. The complex nature of radar's hardware-physical level, e.g., use of high-frequency signals, specialized hardware, fast internal clocks, and protections against signal-noise, have made adversarial interference difficult. So far, only partial success has been shown in constrained environments \cite{komissarov2021spoofing, nashimoto2021low,chauhan2014platform}, with the majority of research being focused on potential, mostly theoretical, countermeasures \cite{katsilieris2013detection, vtc_18}. This is also noted in \cite{chauhan2014platform}, which mentions that \emph{deploying spoofing attacks against commercial automotive radars operating in the mmWave frequency range, poses significant difficulties when compared to evaluations conducted in simulated or constrained \textit{wired} setups}. The two most important highlights of their findings are, first, the difficulty of synchronizing to the victim's carrier frequency (or start frequency), and second, the problem of accounting for, and then minimizing self-interference due to coupling and reflections.

In this work, we extend the state-of-the-art by analyzing the viability of spoofing attacks against truly wireless automotive-grade radars operating at 77GHz. Using COTS hardware in a self-built physical testing framework, we show that such attacks are indeed possible, and provide both theoretical and practical details about how our attacks can be constructed and deployed.

\subsection{Contributions}
We propose (to the best of our knowledge) the first truly wireless attacks against automotive-grade mmWave FMCW radars operating in actively used automotive frequency range of 76-78GHz. Specifically, in this paper we show:
\begin{enumerate}
    \item The first object spoofing attack against an automotive radar which consists of virtual moving objects.
    \item A new practical method of injecting multiple virtual objects at custom angles.
    \item A novel approach to generating and leveraging noise constructively, to remove physical objects and enhance object spoofing attacks.
    \item An in-depth analysis of radar signal processing components and algorithms from an adversarial point of view using real, automotive-grade radar devices deployed in the wild.
\end{enumerate}
In addition to the above contributions, we also review the most efficient countermeasures currently proposed in the literature and place them in the context of our attacks.

A demo of our attack, highlighting multiple moving virtual objects and the constructive noise generation, i.e. (1), (2) and (3) above, is available at: \url{https://mega.nz/file/o58EkKCI\#5paFa3KFLIqn3WNyxHLhobRFg8NUyiwiWm8Lx3rsnbM}.

\section{Background: Automotive radar fundamentals}

In the automotive setting, radar systems are used primarily to identify objects in the environment. The identified objects movements are then tracked and assessed in order to determine the threat to the radar equipped vehicle. Being part of a vehicle's safety system, the accurate identification and constant monitoring of objects are necessary requirements. These two actions are based on three measurements: object distance or range, object velocity and object angle-of-arrival, all with respect to the radar system.

Measurements of range, or distance are done by leveraging three physical properties of electromagnetic energy: physical objects reflect electromagnetic waves, electromagnetic waves or signals travel through air at a constant speed (i.e., speed of light) and finally energy travels through space in a straight line. Pulse radars are designed to measure distance using sequential transmit and listen cycles where waves are emitted in pulses during a transmit cycle, which is immediately followed up by a listen cycle where the radar monitors for incoming signals to detect reflections. The big disadvantage of this approach is that measurements are done in chunks, which lead to potential gaps of knowledge.

\begin{figure}[tb]
    \centering
    \includegraphics[width=240pt,trim={20 20 15 15},clip]{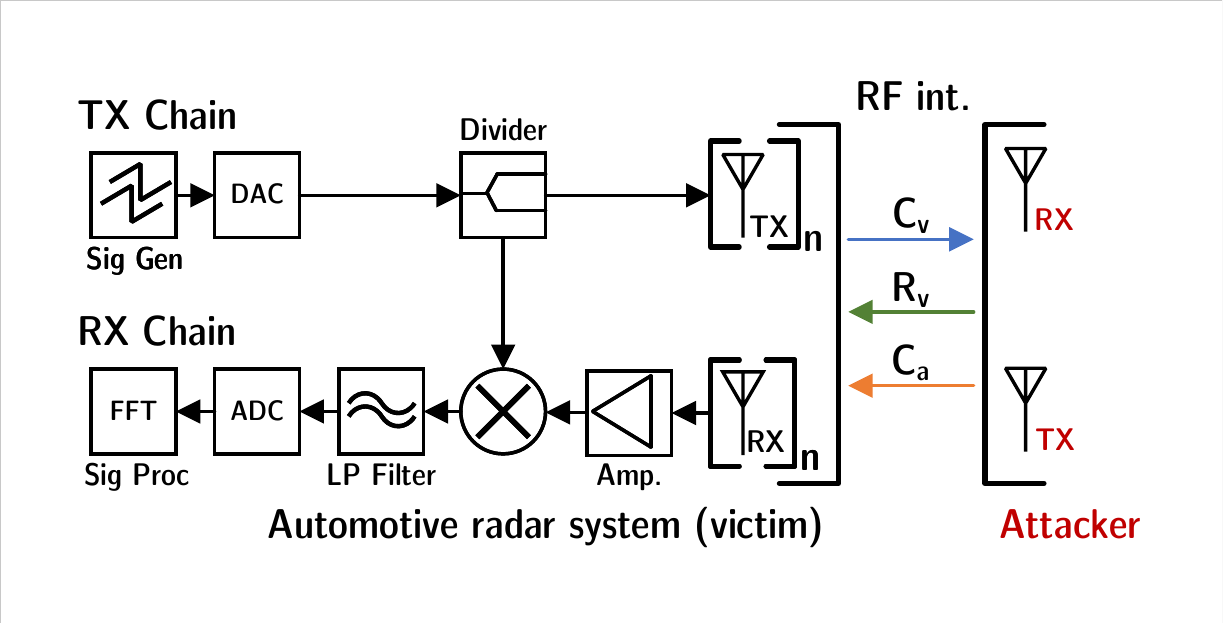}
    \caption{Radar block diagram (left), a minimalistic attacker (right) and the signals between the two (middle).}
    \label{fig:radar_diag}
\end{figure}

Continuous-wave radars (CW radar) address this problem by transmitting continuously, on a known stable frequency, a wave of electromagnetic energy while also listening for waves from any reflecting objects. Moving objects can be detected using the Doppler effect, which causes the received signal to have a different frequency from the transmitted wave. However, the reliance on Doppler-shift detection means that unmodulated continuous wave radars cannot measure distance, only the velocity of moving objects.

CW radars can be further improved by modulating the frequency of their transmitted signal combining the properties of both pulse and CW radars into the frequency-modulated continuous wave radar, enabling both distance and speed measurements as needed in order to distinguish and track objects in an automotive setting. 
Automotive FMCW radar systems operate on high frequencies (e.g., between 76GHz and 81GHz) which enable very precise distance measurements (e.g., $\approx 10^{-2} m$) and also facilitate target measurements in close proximity both to one another and to the radar system itself (i.e., the minimal measured range is proportional to the transmitted wavelength).

In this paper we are looking at several attacks which tamper with these safety critical measurements by generating precise electromagnetic waves that are misinterpreted by the victim radar, therefore in order to understand these it is important to first clarify the radars components that play a role in our attack and how the measurements are performed in the absence of an attacking party. As such, in the following we give details on the building blocks of a radar system followed by an overview on how FMCW radars measure range, velocity and angle-of-arrival of objects.

\subsection{Radar system structure}
In order to attack a radar system, one needs to understand how signals are processed inside a radar ECU. We start with the FMCW radar block diagram shown in Fig.~\ref{fig:radar_diag}-left\footnote{Fig.~\ref{fig:radar_diag}-middle and right sections depict our attacker and the observed RF signals, both detailed in Section~\ref{sec:radar_att_signals}.}. The measurement process starts on the transmit chain, from the signal generator component (e.g., Sig Gen) which produces a sequence of chirps (i.e., chirp frame) according to internally configured parameters such as: chirp bandwidth $b$, chirp duration $t_c$, chirps per frame $N$, and frame duration $t_{f}$. This digital signal is converted to analog and then sent out using a TX antenna array. This outbound analog signal is also internally routed to the receive chain using a power divider in order to provide the TX component of the intermediate frequency (IF) signal.

On the receive chain, reflections are measured by the RX antenna array. The measured signal is amplified and then combined with the internally routed TX signal to construct the IF. After the combination, filtering is performed on the analog IF signal in order to eliminate unwanted noise. Finally, the IF is converted into digital format and fed into a signal processing component which can process this signal and extract the distance, velocity, and angle-of-arrival of the objects in front of the radar using FFT, CFAR and Windowing techniques as described in the following.

\subsection{Object range measurement}
\label{sec:range}

\begin{figure}[tb]
     \centering
     \begin{subfigure}{112pt}
        \centering
        \includegraphics[width=112pt,trim={15 15 15 20},clip]{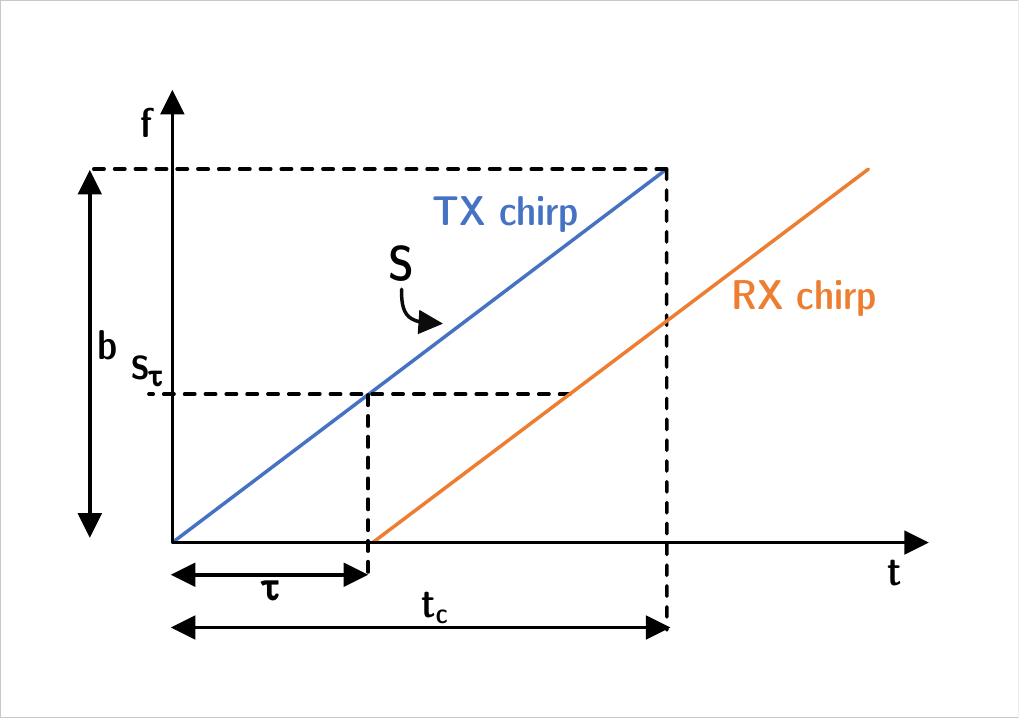}
        \caption{Chirp signal waveform where: TX line represents the transmitted signal shown as a frequency variation over time; RX line as the received signal; $b$ and $S$ are the bandwidth and slope of the TX and RX signals; $\tau$ is the delay time between the TX and the RX chirps and $t_c$ is the chirp duration. $s_\tau$ is the IF, a constant frequency corresponding to a reflected object.}
        \label{fig:chirp}
    \end{subfigure}
    \hfill
    \begin{subfigure}{112pt}
        \centering
        \includegraphics[width=112pt,trim={15 25 15 20},clip]{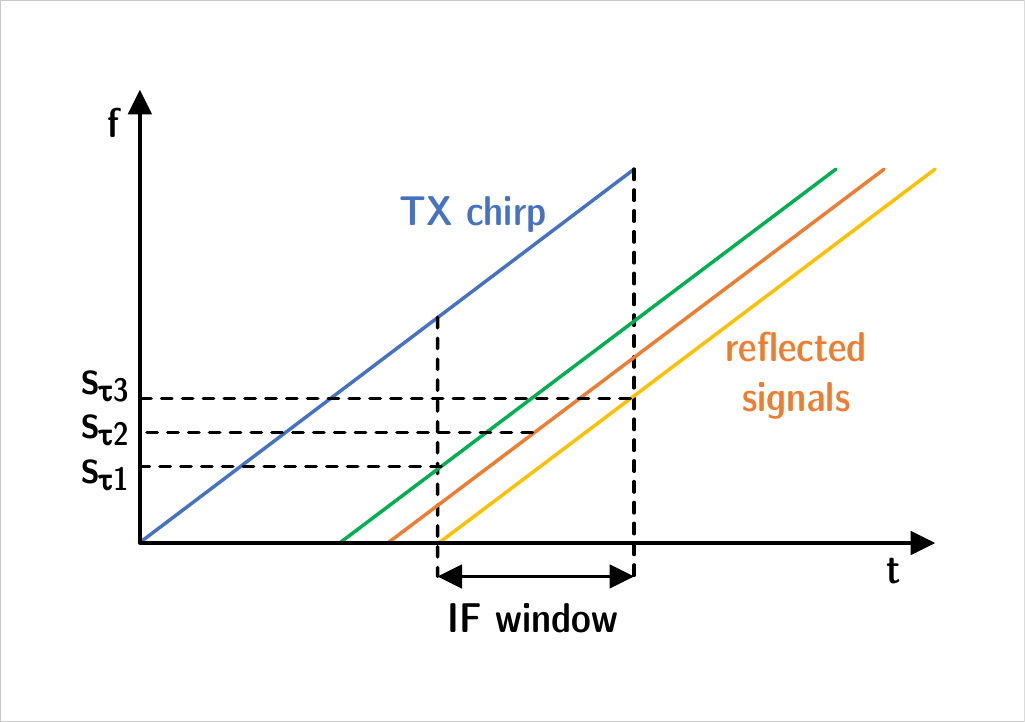}
        \caption{The radar receives a combination of RX reflections corresponding to physical objects. These reflections are combined internally with the TX signal to obtain the intermediate frequency (IF) signal. $s_{\tau1\ldots3}$ are the individual IF tones corresponding to each reflected objects and are extracted using the Fourier transform.}
        \label{fig:multiobj}
    \end{subfigure}
    \caption{Radar chirp parameters}
\end{figure}

The distance measurement is the main method of identifying objects in the environment and, therefore, manipulating this measurement is one of the main components of a successful object spoofing attack. In the following, we give a high level description on how ranges are measured by standard radar system. In Section~\ref{sec:attack}, we build on this and show how attacker generated signals can be constructed in order to trick the victim radar into identifying non-existing, virtual objects.

The measurement process begins with a signal called \textit{chirp} which is transmitted towards a target using a dedicated TX antenna. Modern automotive radars use a saw-tooth waveform chirp that is characterized by a starting frequency $f_s$, a bandwidth $b$, and duration $t_c$. The chirp is also described by a slope $S$ which depends on $b$ and $t_c$ i.e., $b=S\cdot t_c$, which describes the rate at which the chirp ramps up (see Fig.~\ref{fig:chirp}).
Once the chirp is emitted, it travels to the target object, which reflects it back. The reflection signal is received by the radar using a separate RX antenna. Inside the radar system, the TX and RX signals are mixed together. The frequency of the resulting combined signal is called intermediate frequency (IF) and represents the difference between the instantaneous frequency of the TX-chirp and RX-chirp. Any single object in front of the radar will produce an IF signal that has constant frequency tone. The frequency of this tone will be $s_\tau$, and represents the value which allows the computation of the distance to the object $d$, as shown below:

\begin{equation}
    \label{eq:range}
    s_\tau=\frac{S2d}{c_0}; d=\frac{c_0 s_\tau}{2S}=\frac{c_0\tau}{2}
\end{equation}
where $\tau$ is the round-trip duration of the signal and $c_0$ is the speed of light.

The above equation describes the situation when the radar has detected a single object, however FMCW radars can detect multiple objects within one chirp. If multiple objects exist in front of the radar, then each of them will reflect the TX chirp, and the RX chirps received will be delayed proportional to the distance to each of the objects (see Fig.~\ref{fig:multiobj}). The Fourier transform is used to process the IF signal comprised of the TX chirp and the RX reflections to extract the individual frequencies corresponding to each object. The range resolution, or the ability to distinguish between multiple objects, is thus given by the ability to separate the frequencies that make up the IF signal. This is proportional to the bandwidth $b$ of the transmitted chirp signal and is derived from the Fourier transform theory as $d_{res}=\frac{c_0}{2b}$.

\subsection{Object velocity measurement}
\label{sec:velocity}

In addition to range measurements, FMCW radars can also measure relative velocity of objects. This measurement is important because it allows a radar to further distinguish between objects located at a specific distance based on their relative velocities. Our non-synchronized adversary, described in Section~\ref{sec:res_non-sync}, exploits some of the limitations of this measuring technique, in order to generate non-static virtual objects.

Intuitively, the relative velocity of an object can be computed as the distance travelled by the object between the two positions over a known time interval $t_c$. These distances can be measured using the range-FFT described above, using only two chirps separated by the time $t_c$. This type of measurement is, however, not suitable in practice, because the high-frequency velocity measurements that are needed to support real-time tracking of cars, obstacles, etc., are not achievable even with very small values of $t_c$. One important reason for this it that the reflections produced by these closely grouped chirps will have identical tonalities for the IF signal, making velocity measurements impossible using just frequency information. To address this, FMCW radars rely on the phase information contained in the reflected signal. By measuring phase changes between the chirps, small motions of objects can be detected which otherwise would remain undetected by the range-FFT.

\begin{figure}[tb]
    \centering
    \includegraphics[width=240pt,trim={25 20 15 15},clip]{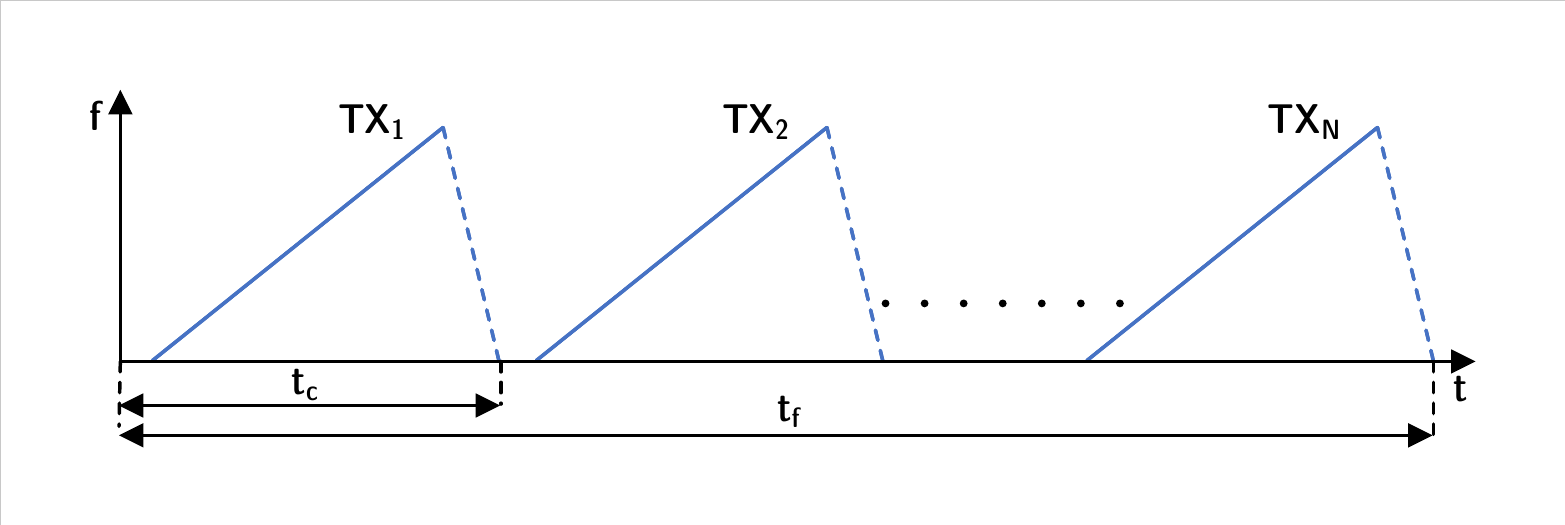}
    \caption{Chirp frame with duration $t_{f}$ comprised of $N$ uniformly distributed identical chirps, each of duration $t_c$. The signal acquisition/sampling for the IF only happens during the frequency ramp-up period (solid blue line) and the ramp-down (dotted blue line) is ignored.}
    \label{fig:frame}
\end{figure}

Velocity measurements are therefore done in several steps. Initially, the radar system transmits and measures a sequence of $N$ equally spaced chirps grouped together in a \textit{chirp frame} as shown in Fig.~\ref{fig:frame}. Then a range-FFT is computed for each of the $N$ chirps inside the frame. The range-FFT will show identical peaks for each chirp in the chirp frame, but each peak will contain multiple phase contributions. A second FFT, called Doppler-FFT, is then used to extract this phase information as a phase difference, $\Delta\phi$. This difference corresponds to the small movements that occurred during the measurement interval and enables the computation of the velocity as:

\begin{equation}
    \label{eq:velocity}
    v=\frac{\lambda\Delta\phi}{4 \pi t_c}
\end{equation}
where $\lambda$ is the wavelength of the reflected signal.

\subsection{Object angle-of-arrival measurement}
\label{sec:aoa}

\begin{figure}[tb]
    \centering
    \includegraphics[width=220pt,trim={15 20 15 15},clip]{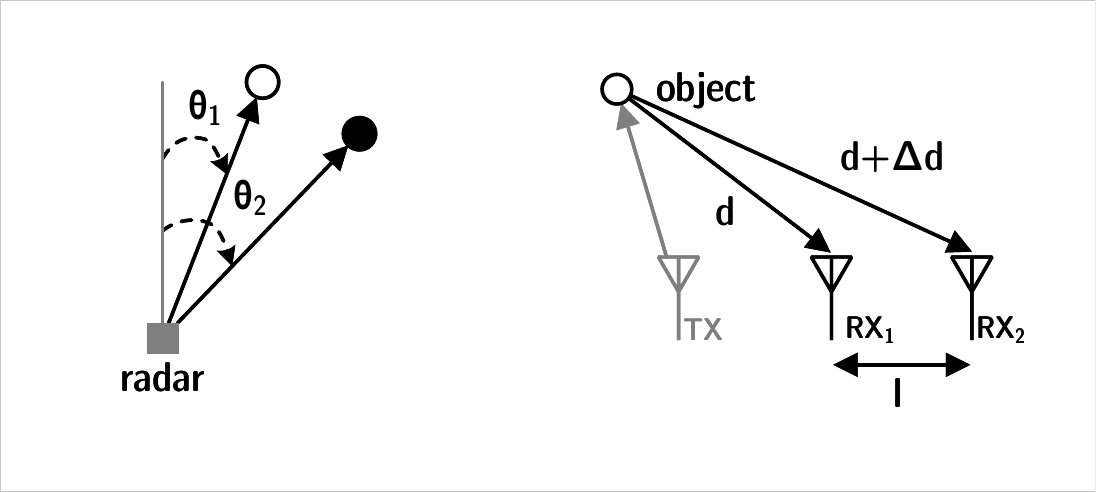}
    \caption{Angle of arrival. Left: radar view of two objects as given by angles  $\theta_{\{1,2\}}$. Right: $\theta$ computation using two simultaneous distance measurements $d$ and $d+\Delta d$ done at antennas $RX_{\{1,2\}}$ respectively, and the fixed distance between them $l$.}
    \label{fig:aoa}
\end{figure}

Lastly, FMCW radars are also able to identify objects based on their angle-of-arrival (AoA) (see Fig.~\ref{fig:aoa}). Our \textit{multiple virtual object generation} attack, shown in Section~\ref{sec:multi_obj}, exploits the internal details of this measurement in order to generate several virtual objects using a single adversarial device. A brief description of how radars compute AoA is presented in the following.

The angle-of-arrival measurement is similar to the \textit{object velocity measurement} in the sense that it leverages the FMCW radars ability to detect small phase changes in the peaks obtained from the range-FFT or Doppler-FFT which correspond to variations in the distance travelled by the reflected signal. However, instead of measuring the phase difference between multiple chirps, angle estimation relies instead on the ability to measure the phase difference between multiple physically separated locations for a single chirp. Therefore, in order to conduct AoA measurements, the radar system needs to be equipped with at least two RX antennas.

For a horizontal plane and two receive antennas i.e., $RX_{\{1,2\}}$ separated by a physical distance $l$, as shown in Fig.~\ref{fig:aoa} the extra distance travelled by the reflected signal, $\Delta d$, to reach $RX_2$ can be computed as $\Delta d=l sin(\theta)$, where $\theta$ is the angle-of-arrival. However, this distance can also be defined as a function of the phase difference $\Delta d=\frac{\lambda\Delta\phi}{2\pi}$. This enables the computation of the AoA from the measured $\Delta\phi$ as:
\begin{equation}
    \label{eq:aoa}
    \theta=arcsin(\frac{\lambda\Delta\phi}{2 \pi l})
\end{equation}

\subsection{Signal processing}
\label{sec:sig_proc}


In radar systems, signal processing algorithms play the crucial role of determining which reflected signals represent noise and which ones represent actual objects by working in conjunction with the measurements types described above. It follows that, in order to successfully generate virtual objects in an adversarial setup, one needs to understand the functionality and inner workings of these algorithms. In the following we present their standard mode of operation and in Section~\ref{sec:noise_removal} we detail how adversarial generated signals can bias the internal state of these algorithms to aid in constructive physical object removal, which complements virtual object generation and identification.

\subsubsection{Constant false alarm rate (CFAR) algorithms}
\label{sec:cfar}

\begin{figure}[t]
    \centering
    \includegraphics[width=240pt,trim={0 0 0 0},clip]{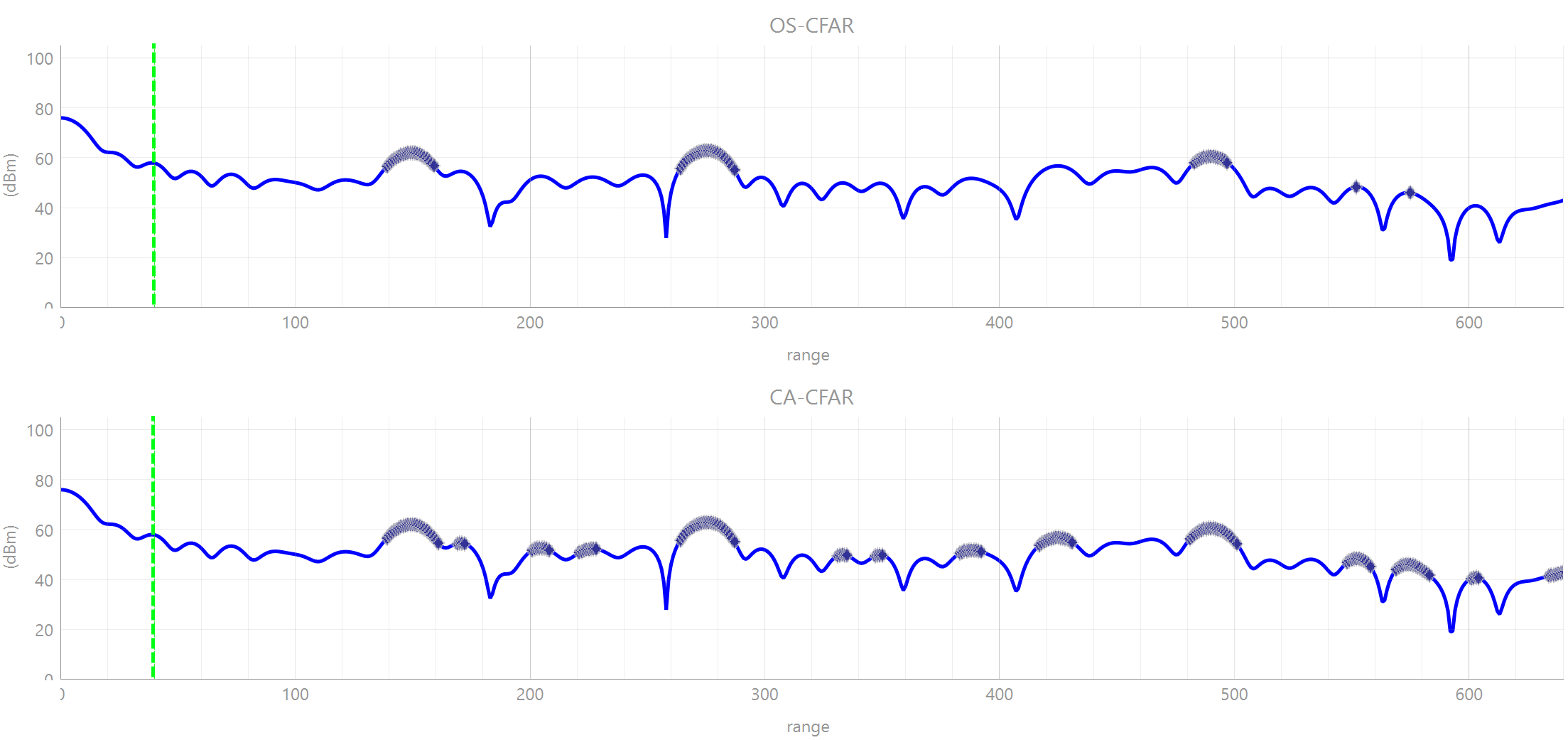}
    \caption{CFAR range-FFT plot. Highlighted peaks represent the physical objects identified by OS-CFAR (top) and CA-CFAR (bottom). Range (cm) is shown on the 0x axis and signal strength of the reflections (dBm) is shown on the 0y axis. The green vertical asymptote (at $x=40$) represents the minimum measurement range imposed by the CFAR window.}
    \label{fig:cfar_graph}
\end{figure}

One important challenge in object detection is to determine whether the FFT spectrum peaks produced by the radar front-end correspond to an objects or not. Threshold values can be used to distinguish between noise and objects, however computing these values statically, in ideal situations, leads to unpredictable results such as either too many false alarms if the threshold is too low, or poor object detection if the threshold is too high. The constant false alarm rate (CFAR) algorithms mitigate the issue by dynamically computing a noise floor that is adapted to the environment measured. This ensures that the number of false alarms does not depend on the noise power level, and facilitate detection rate adjustments such that both safety and environment noise constraints are taken into account. In the following, we describe the most common CFAR algorithms used by the automotive radars to estimate the noise floor: cell averaging (CA)-CFAR and ordered static (OS)-CFAR. The reference output of these can be seen in Fig.~\ref{fig:cfar_graph}. We have used both algorithms in the evaluation of our attacks.

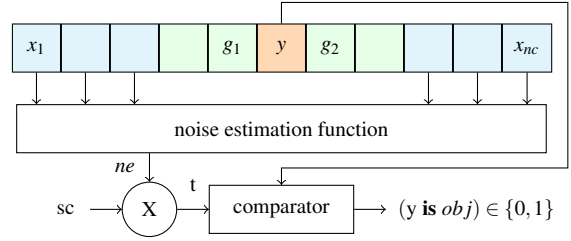
\begin{figure}[t!]
	\begin{center}
		\begin{tikzpicture}[every node/.style={node distance=0cm,rectangle,draw,minimum height=.8cm,minimum width=.8cm,inner sep=.2cm}, scale=0.8, transform shape]
    		\node [fill=cyan!10](x1) {$x_{1}$};		
    		\node [fill=cyan!10,right=0 of x1] (x2) {};
    		\node [fill=cyan!10,right=of x2] (x3) {};
    		\node [fill=green!10,right=of x3] (g1) {};
    		\node [fill=green!10,right=of g1] (g2) {$g_{1}$};
    		\node [fill=orange!30,right=of g2] (y) {$y$};
    		\node [fill=green!10,right=of y] (g3) {$g_{2}$};
    		\node [fill=green!10,right=of g3] (g4) {};
    		\node [fill=cyan!10,right=of g4] (x6) {};
    		\node [fill=cyan!10,right=of x6] (x7) {};
    		\node [fill=cyan!10,right=of x7] (xn) {$x_{nc}$};
    		
    		\node [below=15pt of y,minimum width=11*.8cm] (avg) {noise estimation function};
    		
    		\draw [->] (x1) -- (avg.north -| x1);
    		\draw [->] (x2) -- (avg.north -| x2);
    		\draw [->] (x3) -- (avg.north -| x3);
    		\draw [->] (x6) -- (avg.north -| x6);
    		\draw [->] (x7) -- (avg.north -| x7);
    		\draw [->] (xn) -- (avg.north -| xn);
    		
    		\node [below=15pt of avg,minimum width=3*.8cm] (cmp) {comparator};
    		
    		\draw [->] (y.north) -| ([yshift=10pt] y.north) -| ([xshift=10pt] avg.east) |- ([yshift=8pt] cmp.north) -- (cmp.north);
    				
    		\node [left=15pt of cmp,draw,circle] (mul) {X};
    		\draw [->] (mul.east) -- (cmp.west) node[draw=none,midway,above]{t};
    		\draw [->] (avg.south -| mul.north) -- (mul.north) node[draw=none,midway,left]{$ne$};
    		
    		\node[left=15pt of mul, draw=none] (s) {sc};
    		\draw [->] (s.east) -- (mul.west);
    		
    		\node[right=15pt of cmp, draw=none] (bool) {$($y \textbf{is} $obj) \in \{0,1\}$};
    		\draw [->] (cmp.east) -- (bool.west);
    		
		\end{tikzpicture}
	\end{center}
	\caption{CFAR block diagram for evaluating a range cell $y$. First, a noise estimation function is applied on $y$'s neighboring cells $x_{1..nc}$. Two or more guard cells (e.g., $g_{1,2}$) are used to isolate $y$, the target cell, from the rest of the values in the noise reference window. The noise estimation function's output, $ne$, is scaled using a precomputed value $sc$ to obtain the noise threshold $t$ of the window. Finally, $y$ is compared to the noise threshold to determine whether it represents an object or not.}
	\label{fig:cfar}
\end{figure}

\noindent\textit{CA-CFAR}. The cell-averaging CFAR algorithm begins by splitting the raw measurements from the FMCW radar system into range cells. A sliding-window technique is then used to determine the noise threshold for each tested cell by averaging over $nc$ neighboring cells, $x_{1..nc}$, as shown in Fig.~\ref{fig:cfar}. Cells are derived from the frequency bins generated by the FFT from the raw signal received by the radar. Each $nc$ value window is made up of three types of cells: the \textit{test cell} which is selected to be in the middle of the interval, the noise \textit{neighbor cells} which are split evenly to the left and to the right of the target cell, and finally guarding cells which are cells adjacent to the test cell but which are not used by the noise estimation function. Guarding cells are especially important because when measuring physical objects, FFT peaks often spread over several frequency bins, extending into neighboring range cells. Another important aspect to note is that the minimum measurement range of a radar using CFAR is defined by the size of the window. This is because the first test cell that can be evaluated needs to be located at least $\frac{nc}{2}+1$ range cells away from the radar front. As such, for a reference window of $nc$ values, the basis for the noise estimation function used by CA-CFAR is a \textit{mean value function}: $ne=\frac{1}{nc}\sum_{i=1}^{nc}{x_{i}}$.

Adaptive threshold radar systems, however, are characterized by \textit{fixed probability of a false alarm} $P_{FA}$ values and window size values $nc$. Thus, in order to use the result from the averaging process $ne$ to compute the noise threshold value $t$, and determine whether the tested cell is an object or not a scaling factor value, $sc$, is used. This value is computed as: $sc=nc(P_{FA}^{-\frac{1}{nc}}-1)$ using statistical properties of the probability density functions associated to the range cells after square law is applied \cite{sqarelaw}, and the probability density function of the test cell $y$. Complete details for deriving the equation are available in \cite{katzlberger2018object}. 

After computing threshold values $t$ for each reference window, the CA-CFAR algorithm returns a Boolean array which states if the test range cell $y$ has exceeded the adaptive noise threshold computed or not.

\noindent\textit{OS-CFAR}. The ordered-static CFAR algorithm has the same design as CA-CFAR (see Fig.~\ref{fig:cfar}). The only significant difference comes from the noise estimation function used. Where CA-CFAR used a mean value function to estimate the noise in a reference window, the OS-CFAR algorithm uses an ordered-static function where noise is estimated by looking at the $k^{th}$ order value in a set comprised of the values associated to the neighboring cells, sorted in ascended order. As such, if multiple objects are located in the reference window, they do not affect the detection in the cell under test $y$. According to \cite{oscfar} the probability of false alarm for an OS-CFAR algorithm evaluated on a value $k$ can be reduced to:
\[
    P_{FA}=k\binom{nc}{k}\frac{(k-1)!(sc+nc-k)!}{(sc+nc)!}
\]
By solving the above equation the scaling factor, $sc$, can be obtained for suitable fixed values of $P_{FA}$. We discuss the specific parameters we used for each CFAR algorithm in order to verify our attacks in Section~\ref{sec:sw_conf}.

\subsubsection{Windowing algorithms}
Weighting functions, also known as windowing functions, are algorithms used to address issues related to spectral leakage. As stated previously, the FFT of a reflected signal tends to spread into adjacent frequency bins, causing the phenomenon known as spectral leakage. Ignoring the spectral leakage can result in the non-detection of target signals and therefore of objects. CFAR algorithms use guarding cells to mitigate these effects, however windowing functions are a more efficient, complementary method of dealing with the issue.

Windowing functions are applied to the captured signal before it passes to the FFT stage, thus before frequency bins are extracted, behaving more or less like an electronic signal filter. For distances smaller than 31 nautical-miles (e.g. 57~km) there is no significant difference between windowing algorithms, however Hann still remains one of the recommended windowing functions for reducing spectral leakage and enhancing frequency resolution \cite{windowing}. The  radars we use have hardware level implementations of the Hann windowing algorithm, and thus we evaluate our attacks against a victim which filter signals using this technique in order to enhance the practicality of our evaluation.

\section{Attacking FMCW radars}
\label{sec:attack}



\subsection{Threat model}
Our threat model has been introduced previously in Fig.~\ref{fig:radar_diag}. In our model, we only have two requirements. First, the adversary needs access to a radio interface such as a secondary radar device. These are usually available as COTS devices \cite{inras_rbk2} or can be retrofitted from existing automotive radar ECUs. Alternatively, a specialized SDR capable of generating chirps in the victim's frequency range (e.g., 76-78GHz) could be used. As far as we are aware such SDRs do not exist, which, we speculate, is why truly wireless attacks have not been demonstrated up until now. Second, the adversary also requires some generic information about the implementation of the victim's radar ECU, in order to narrow down the suitable attack parameters. Fortunately, obtaining these constant parameters is a one-off effort which can easily be accomplished by surveying commercial radar specs such as ECU user manuals or promotional materials. Furthermore, we have observed that these are common across radar units \cite{conti1,conti2}.

As such, the first step of the attack consists of obtaining the victim's chirp parameters, i.e., the starting frequency ($f_s$), the bandwidth $b$ and the chirp duration $t_c$, along with the frame length $N$ and the frame duration $t_{f}$.
At this point, based on how these parameters are configured, we distinguish two types of adversaries: \textit{synchronized} and \textit{non-synchronized}. The \textit{synchronized adversary} is stronger, and leverages prior synchronization with the victim before deploying the attack, as well as constant adjustment feedback throughout the attack. The weaker \textit{non-synchronized adversary} does not require any type of synchronization, and only leverages knowledge of the initial parameters.

Following the initial parameter learning and configuration, the attack continues with the main objective: generating virtual objects using adversary generated signals. We strictly require that these objects must be indistinguishable from actual physical object reflections, when observed by the victim radar. In order to achieve this, attacker generated signals need to be able to reach the signal processing component (e.g., Sig Proc) of the receive chain of the victim (see Fig.~\ref{fig:radar_diag}) while still possessing all the characteristics of an actual reflection. The attack is considered successful if the victim is able to infer the existence of objects by computing range, velocity or AoA using these signals.

\subsection{Preliminary: adversarial signals}
\label{sec:radar_att_signals}

From an adversary point of view, the first challenge in generating virtual objects, is to influence the intermediate frequency signal of the victim. This, however, is a difficult undertaking given that the adversary has only partial control of the signals observed by the victim and, therefore, limited capability to influence them. In order to get a better picture, briefly, let us consider the following signals observable at the radar's RF interface (also shown in Fig.~\ref{fig:radar_diag}): $C_v$ the victim's transmitted chirps, $R_v$ the legitimate reflections of physical objects, and $C_a$ the adversary transmitted chirps. Then, by definition, when the victim's transmitted signal is $TX_v = C_v$, its received signal is $RX_v = R_v \otimes C_a$. The victim's intermediate frequency signal $IF_v$ is therefore: $$ IF_v = C_v \otimes ( R_v \otimes C_a ) $$
However, out of these three signals comprising $IF_v$, the adversary can only affect the victim's intermediate frequency, through its generated signals $C_a$. Victim transmitted chirps $C_v$ can only be leveraged to gain information about the internal parameters of the victim radar and, most importantly, the adversary has no ability to measure or observe $R_v$, the physical object reflections as seen from the victim's point of view. This represents one of the biggest challenges in deploying object-spoofing attacks in truly wireless environments \cite{chauhan2014platform}. In Section~\ref{sec:realworldexp} we detail our method to overcome this.

\subsubsection{Adversarial parameters for object generation}
Successful virtual object generation requires an adversary that is able to control the following three components: \textit{signal shape}, \textit{delay} and \textit{power}. The \textit{signal shape} aspect refers to the adversary's ability to produce signals that are identical to signals produced by physical objects when they are illuminated by the victim transmitted signals. An adversary with knowledge of the chirp parameters of the victim will be able to produce valid signals, given that the signals reflected by real objects will be identical to those with which they were illuminated.

\begin{figure}[tb]
    \centering
    \includegraphics[width=240pt,trim={20 15 15 15},clip]{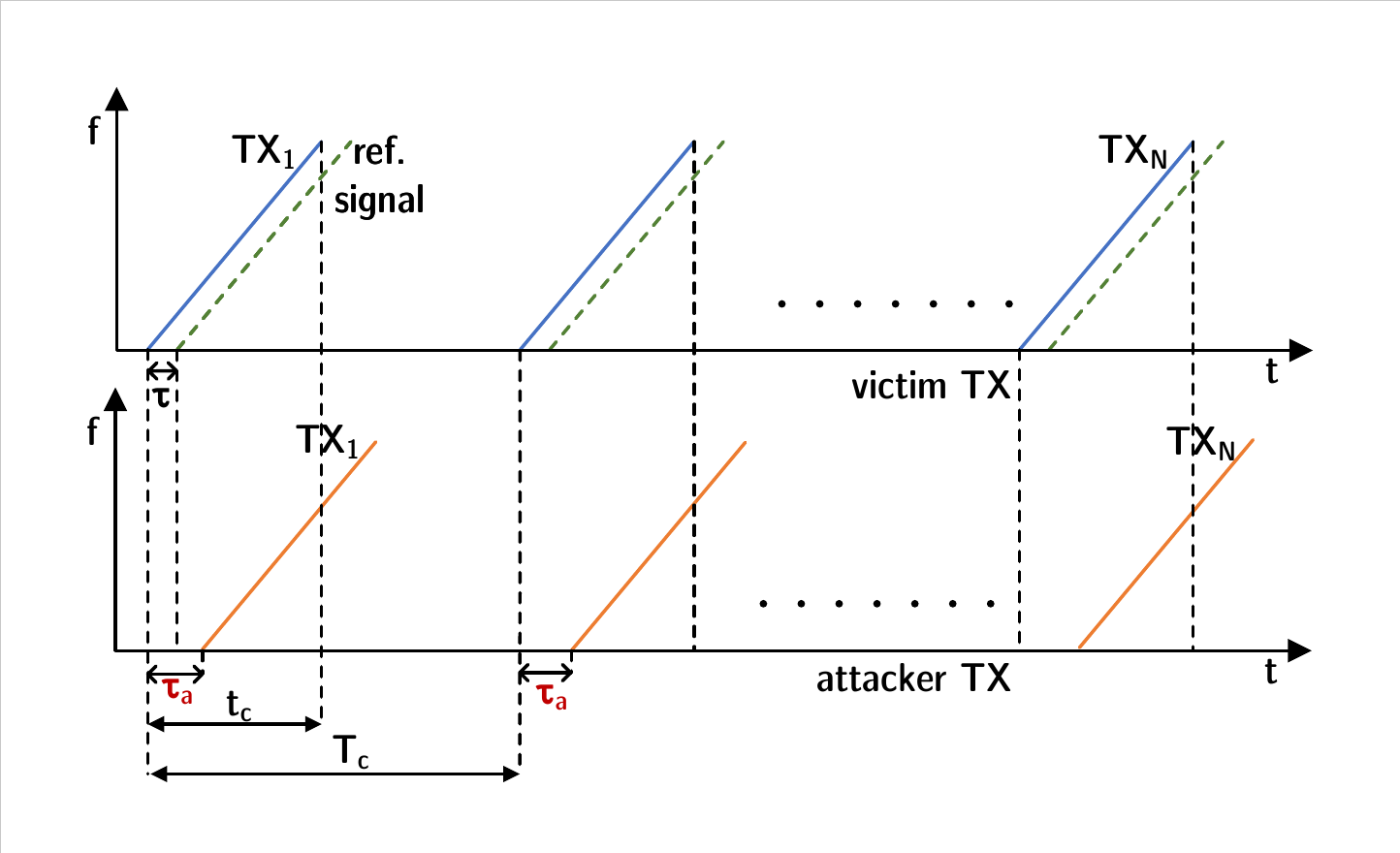}
    \caption{Synchronized adversarial scenario. Top represents a standard sequence of chirps (i.e., frame) sent by a victim radar device. The expected reflected signal is shown as a dotted green line. Bottom represents the attacking radar's chirps, which are identical to the victim radar's chirps, but at a slight offset. $t_{c}$ is the chirp duration during which the victim radar samples (i.e., listens to) the RF signals. $T_c$ is the chirp period of the victim. $\tau$ is the time delay of a signal reflected by a physical object. $\tau_{a}$ is an attacker controlled delay which, if chosen appropriately, results in aligned signals which are valid reflections when received by the victim radar.}
    \label{fig:sync_att_time}
\end{figure}

The \textit{delay} aspect refers to the adversary's ability to control the time delay of its injected signals or chirps with respect to the victim's transmitted chirps. Given that distance to objects is derived based on measurements of this delay (see Section~\ref{sec:range}) it represents the main component in virtual object generation. Controlling delay however is not trivial for several reasons. First, a typical radar system will only listen to signals (i.e., perform data acquisition) while also transmitting them. This is required in order to generate the IF signal, and it is fundamental to the functionality of FMCW radars. A successful adversary will need to be sufficiently synchronized to the victim in order to be able to inject any signals. Second, radar transmitted signals are not individual chirps but rather chirp frames, thus requiring the adversary to synchronize with multiple chirps consistently. And finally, due to reasons related to energy consumption, the transmission (and listening) time within a frame represents $\approx1\%$ of the total frame duration\footnote{A typical frame with duration $t_{frame}=200ms$ consisting of $50$ chirps with duration $t_c=51.2\mu s$ has an active transmission time of $1.28\%$ or $2.56ms$.} which adds further, even higher constraints, on the precision requirements of the adversary.

Lastly, the \textit{power} aspect refers to the adversary's ability to transmit signals with intensities/amplitudes comparable to those generated by the illuminated objects. The intensity of this signal, considering its path as starting from a source radar, going to an object and then getting back to the radar, is inversely proportional to the square of the distance, according to the inverse square law, but applied twice as the signal travels in both directions, namely $I \propto \frac{1}{d^4}$. 
This, however, is not trivial as the adversarial injected signals need to match this intensity in order to circumvent noise filtering hardware (e.g., electronic low-pass filters) and software techniques (e.g., CFAR) whilst travelling the distance only once.

In the following, we describe, first, the classic adversarial scenario which relies on a synchronized adversary. Then we introduce our new non-synchronized adversary. Finally, we show two new capabilities of the non-synchronized adversary that were previously not known: multi-object generation, and a constructive noise generation method that removes physical objects and highlights virtual ones.


\begin{figure}[tb]
    \centering
    \includegraphics[width=240pt,trim={20 15 15 15},clip]{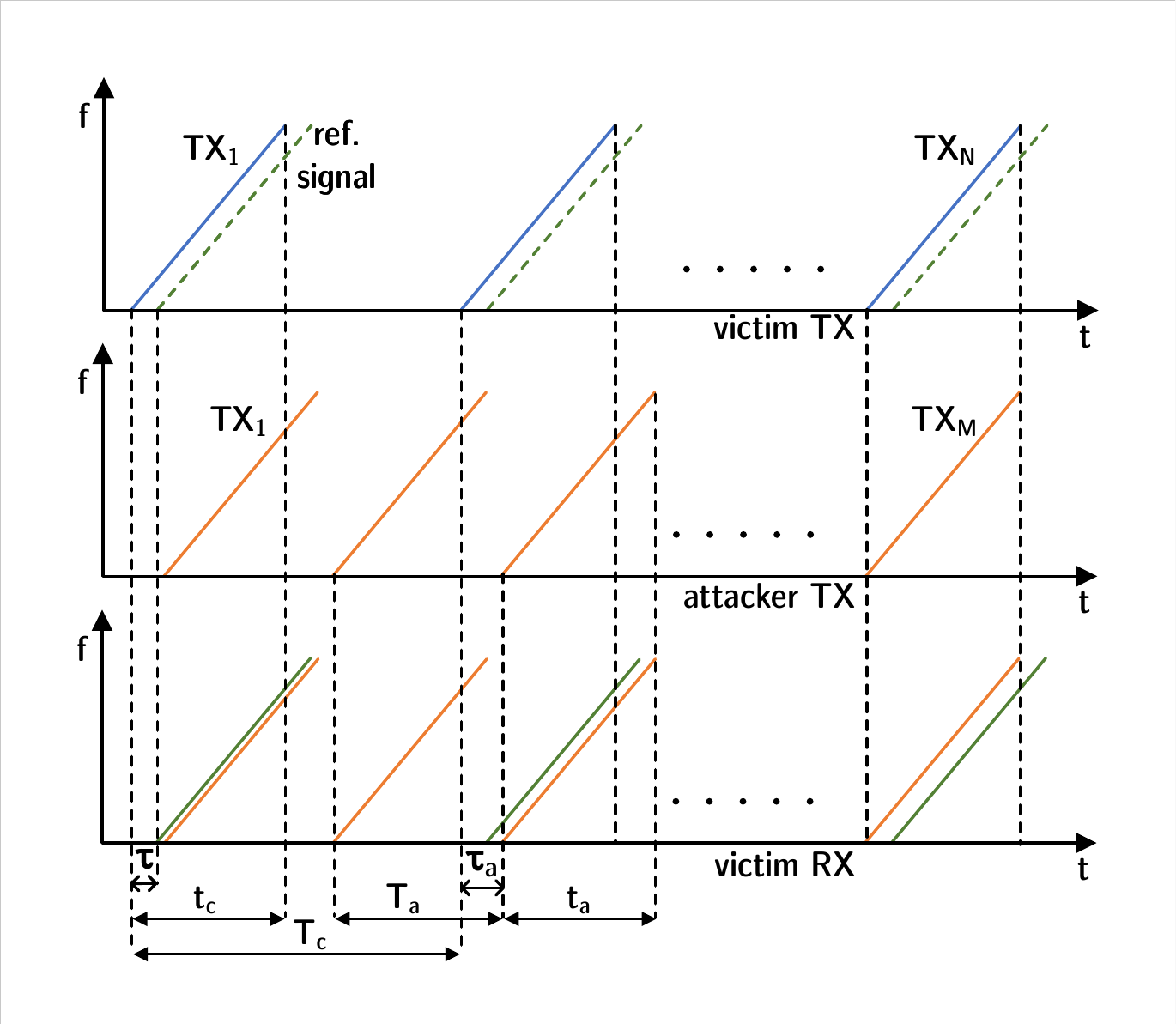}
    \caption{Non-synchronized adversarial scenario. Top represents a standard sequence of chirps (blue) and expected reflected signal (green) as perceived by the victim radar device. Middle represents the attacking radar's chirps. Bottom shows the actual signals received by the victim radar: a combination of legitimate reflected chirps (green) and the injected chirps (orange). Notation as in Fig.~\ref{fig:sync_att_time} except for: (1) the chirp period of the attacker $T_a$ is different from the chirp period of the victim $T_c$ and (2) the attacker transmits more chirps per frame than the victim.}
    \label{fig:att_inj}
\end{figure}

\subsection{Baseline scenario: synchronized adversary}
\label{sec:sync_def}

Intuitively, when thinking of radar spoofing attacks, the ones expected to have the highest chances to succeed are those which use a \textit{synchronized adversarial model}. We speculate that this model has been primarily chosen based on its success against other automotive sensors such as LiDAR \cite{shin2017illusion,petit2015remote,sun2020towards,lidar_attack1,lidar_attack2}. Due to the increased complexity of radar sensors (as compared to LiDAR and camera systems), however, only partial success has been demonstrated in constrained, simulated environments \cite{komissarov2021spoofing, nashimoto2021low}. 

We begin with the analysis of the synchronized adversary due to its familiarity and simplicity. We evaluate it later in Section~\ref{sec:res_sync} in order to establish an attack baseline. Our main attacker, however, is the non-synchronized attacker, introduced in Section~\ref{sec:theo_nonsync} which is a weaker, more flexible attacker which does not require synchronization and achieves significantly better results.

As such, in the synchronized adversarial scenario, the adversary deploys its attack in two stages. First, the adversary attempts to synchronize with a victim through passive listening, and learn the period of the victim transmitted chirps, $T_c$. If this can be accomplished successfully, then the attacker continues with the second stage, in which it attempts to maintain chirp-level and/or frame-level synchronization in order to correctly align its self-generated signals to the victim's and produce valid reflections. The relationship between the victim and the attacker radars is shown in Fig.~\ref{fig:sync_att_time} and works as follows. After the adversary learns the victim's chirp period $T_c$, it will compute a fixed time offset $\tau_a$ calculated based on the victim's chirp start time. The attacker then starts transmitting chirps identical to the victim's at $T_c+\tau_a$ intervals. If chirp-level synchronization is maintained, and the value of $\tau_a$ is smaller than the listening window of the victim ($t_c$) then the chirps generated by the attacker will be included in the $IF_v$ of the victim. This will result in an attacker controlled virtual object whose location can be computed using eq.~\ref{eq:range} as: $d_{vo}=\frac{c_0\tau_a}{2}$.

\subsection{Realistic scenario: non-synchronized adversary}
\label{sec:theo_nonsync}

In the non-synchronized adversarial scenario, we are analyzing a realistic adversary which does not require any form of feedback from the victim radar and is completely detached from the victim, both physically and logically. We discuss our results for this adversary in Section~\ref{sec:realworldexp}. We consider this to be a much more realistic adversary in the real world, when compared to the synchronized one, because maintaining any level of synchronization between the victim and the attacker is extremely difficult in practice, mainly due to physical constraints and issues related to hardware differences between the victim and the attacker. 

Thus, a non-synchronized adversarial radar is required to generate valid chirps (i.e., chirps that are interpreted as reflections by the victim), without accurately matching the period of these chirps to the victim. A summary of this scenario is presented in Fig.~\ref{fig:att_inj}. The main difference between the synchronized and non-synchronized attacker scenarios is as follows. In the non-synchronized version, the adversary does not generate its chirps at fixed delays computed as an offset from the victim's transmitted signals. Instead, the attacker computes an independent chirp period $T_a$ that is smaller than the victim's. By carefully adjusting the ratio between $T_a$ and the victim's period $T_c$, the adversary can achieve a high degree of overlap with the listening periods $t_c$ of the victim, and thus influence its intermediate frequency signal. Even though $\tau_a$ is not constant between chirps from the victim's point of view, in this attack, the adversary can still leverage this variability and generate moving virtual objects. This objects will have a computable speed equal to $v_{vo}=\frac{\Delta d_{vo}}{T_c}$, where $\Delta d_{vo}$ is the distance travelled by a virtual object between two separate chirps, where the attacker injected signal is at offsets $\tau'_{a}$ and $\tau''_{a}$. Our experimental results related to this attacker are shown in Section~\ref{sec:res_non-sync}.

\subsection{Multiple virtual object generation}
\label{sec:multi_obj}

The \textit{multiple virtual object generation} scenario is a complementary attack to the above and can be employed both by the synchronized and the non-synchronized adversaries. Normally distance to a physical object, e.g., on the $Ox$ axis, can be measured by extracting frequency tones from reflected signals as shown in Section~\ref{sec:range}. Briefly, a radar unit processes its intermediate-frequency signal using its RX chain, and extracts frequency tones for each physical or virtual object. These tones are then used to determine the distance to the object approximate to a given resolution, also known as a range bin. However, when physical objects are located in close proximity to one another, the frequency tones are no longer enough to distinguish between individual objects and end up being grouped together in the same range bin. In Section~\ref{sec:aoa} we explain that it is still possible to distinguish between multiple objects in the same range bin by measuring changes to the phase of the signal.

\begin{figure}[tb]
    \centering
    \includegraphics[width=175pt,trim={20 15 15 15},clip]{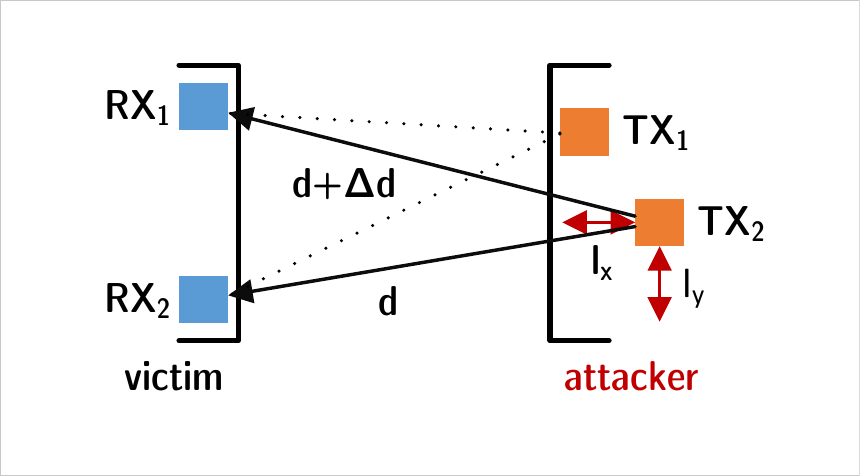}
    \caption{An attacker can generate one virtual object per TX antenna. By adjusting the physical position of its antennas relative to one another, e.g., $(l_x,l_y)$, the attacker can influence the victim perceived spatial position of the virtual object generated. The distance between antennas, on the $l_y$ direction, is usually fixed.}
    \label{fig:multi_obj}
\end{figure}

In this new attack scenario, the adversary leverages the above properties of radars systems to generate multiple objects. This plays an important role as it can be used by the adversary to generate objects that are placed at an angle offset compared to the position of the transmitted antenna, greatly increasing the capabilities of the attacker. An outline of our attack is presented in Fig.~\ref{fig:multi_obj}. The attack assumes a standard victim radar equipped with multiple receive antennas (RX) which enable angle-of-arrival measurements. On the attacker side, the adversary is equipped with two or more transmit antennas (TX) which are all generating identical chirps. The attack consists of physically altering the distance between these in a controlled manner for specific lengths, e.g., $(l_x,l_y)$ along the $Ox$ and $Oy$ axis. Given the small relative distances of $l_x$ and $l_y$ the antenna movement is not sufficient to alter the frequency tone produced by each one, however these movements are enough to affect the phases of the attack signals generated. The effects observed by the victim, regardless of the direction of translation performed by the attacker, will be in the phase spectrum. More specifically, the victim will perceive a cluster of objects, one for each of the attacker's TX antennas, all located at the same range from the victim (i.e., all objects are located in the same range bin), but at different angles as given by the phase alterations induced by $(l_x,l_y)$ translations.

\subsection{Constructive physical object removal through noise}
\label{sec:noise_removal}

\begin{figure}[t]
    \centering
    \includegraphics[width=240pt,trim={0 0 0 0},clip]{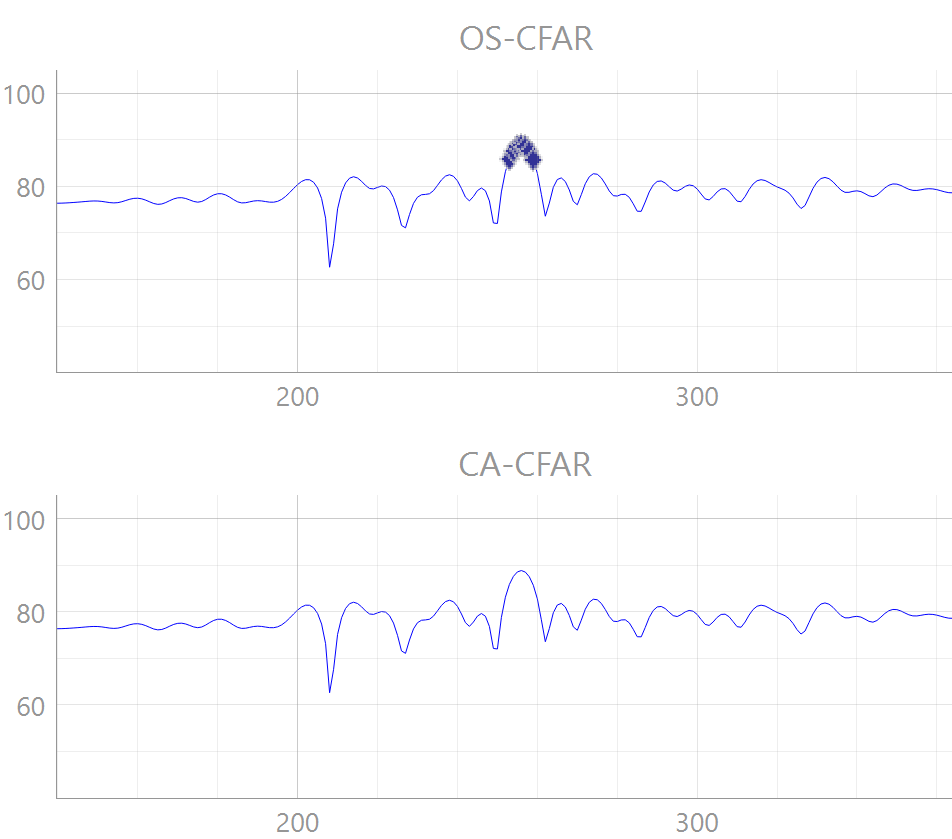}
    \caption{Constructive object removal through noise. The signal-noise is raised to 80dBm. This obfuscates all real-world objects that would have been identified by the CA-CFAR (bottom). The OS-CFAR (top) can still identify one real-world object, a perfect target (e.g., corner reflector target). Range (cm) is shown on the 0x axis, and signal strength of the reflections (dBm) is shown on the 0y axis.}
    \label{noise_fft}
\end{figure}

The constructive physical object removal is our second extended attack which shows for the first time that noise can be leveraged to remove physical objects and highlight virtual objects, going further than the traditional destructive interference that usually just blinds the victim (Fig.~\ref{noise_fft}).

One of the challenges in spoofing radar objects is related to the lack of adversary's control over signal reflected by physical objects present in the victim's environment. Noise insertion has always been viewed as a low-hanging fruit type of attack, where the adversary saturates the victim's bandwidth with high intensity signals in order to blind the victim and hinder detection. However, modern FMCW radars are able to keep track of hundreds of objects and have built methods for detecting, and possibly mitigating, these types of attacks. As far as we have observed, mitigations consist mainly of electronic components, such as band-filters and amplifiers, which ensure that only the suitable frequencies used to generate the chirp are processed by the receive chain. Furthermore, CFAR and Windowing algorithms (see Section~\ref{sec:sig_proc}) complement the electronic protections by enabling dynamic noise adjustment and rejection of outlying signals. Nonetheless, efficient versions for noise attacks have been known in military setups for quite some time \cite{ewf}, and have been demonstrated in realistic scenarios such as \cite{electronics9040573}. However, although these attacks are able to prevent radar operability, they are still detectable through noise level measurements across the radar's observed spectrum.

The goals of the attacker proposed here, however, are different to those highlighted above. First, our adversary aims to generate noise that can bypass the hardware filtering of the victim and is of a suitable intensity such that the attack remains \emph{undetectable}. This is a significant challenge because it requires a significant reduction and dynamic adjustments to the power output of its TX signals in order to match the victim's expected signal intensity levels; and specific adjustments to the ways in which noise is generated such that attack signals generated are not rejected by the CFAR and Windowing filtering algorithms. 

The second goal of our attacker is to constructively leverage the noise generated to aid the detection of virtual objects. This task is accomplished in the following two steps. First, existing physical objects are removed from the victim radar's view by carefully manipulating the inputs to the CFAR and Windowing algorithms. This step effectively raises the noise floor just enough to establish a new environment baseline, but still low enough as to not trigger any for of rejection. This is achieved by the attacker through of its TX signal power levels, mainly by adjusting distance to the victim and/or by modifying antenna gains, and through careful signal alignment such as the one described in for our synchronized and non-synchronized adversaries above. Virtual object highlighting can then be performed by overlapping the TX signal corresponding to the generated virtual objects on top of this noise at the appropriate intervals. We detail our results for this attack in Section~\ref{sec:res_non-sync}.

\section{Experimental setup and results}
\label{sec:realworldexp}

Our experimental setup consists of two identical INRAS Radarbook 2 automotive-grade radars used in the automotive industry as reference radars for calibrating and testing ADAS systems \cite{rhode_sch}, with plans to also be used in the aeronautical industry for electric airplane guidance \cite{jobi}. 

Each device consists of on an Arria 10 ARM System-On-Chip board equipped with a 77-GHz FMCW Frontend \cite{inras_rbk2}. The 77-GHz FMCW Frontend is implemented on an FPGA and supports 16 RX channels and 2 TX channels. The unit is capable of computing range-Doppler maps with up to 2048 range bins and 512 velocity bins, and has built-in hardware support for an FFT processing framework, as well as an imaging and synthetic-aperture radar (SAR) processing frameworks used to create two-dimensional images or three-dimensional reconstructions of objects. 

\begin{figure}[t]
    \centering
    \includegraphics[width=240pt,trim={20 20 20 20},clip]{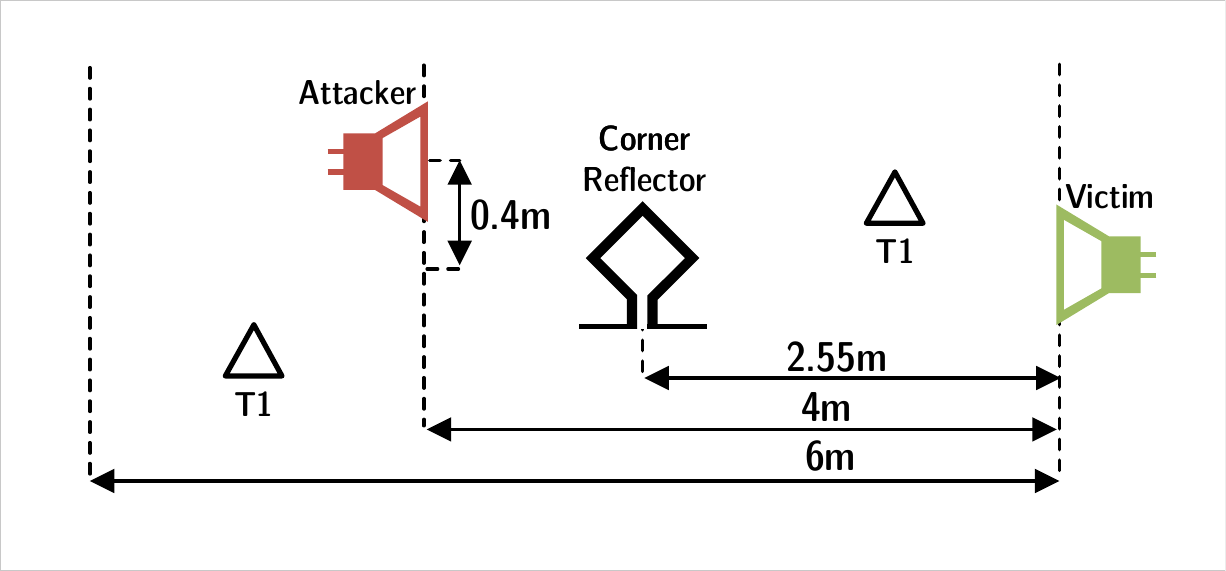}
    \caption{Main physical setup. Two radar units are positioned facing each other at an orthogonal distance of 4m. The attacker is offset from the victim with 0.4m. The victim's view is restricted to a measurement distance of 6m. A control target, in the form of a corner reflector, is positioned 2.55m from the victim, facing it. T1 and T2 are objects used for the evaluation of our constructive physical object removal.}
    \label{fig:physical_setup}
\end{figure}

\subsection{Physical setup}
\label{sec:phy_conf}
Our physical setup is outlined in Fig.~\ref{fig:physical_setup}. This was used to test both the wired synchronized attacker and the non-synchronized wireless one. We intentionally chose a small and fixed measurement distance for benchmarking between the victim radar and the attacker, i.e., $4m$, to determine the impact of the attacker signal intensity on the victim as shown in Fig.~\ref{fig:physical_setup}. It is important to note that in the case of the wireless attack, this distance can easily be adjusted to increase or decrease the intensity of the attacker's signal. Furthermore, we have observed that larger distances (e.g. $10m$) greatly improve the viability of the attack by allowing better control of the injected signal's power, which directly affects the victim perceived noise.

\subsection{Software configuration}
\label{sec:sw_conf}

For our real-world experiments and benchmarking, we have \textit{additionally} implemented CFAR, and windowing algorithms in software. Due to the high computational requirements associated with live signal processing, we were restricted to the built-in hardware FFT functionality when computing range, velocity and AoA bins. When multiple algorithm versions were available, we were able to replicate the results both using the hardware implementation, as well as our own software based analogues. The specific numerical parameters detailed in the following were set to values commonly used by automotive radar manufacturers, and were obtained by surveying commercial radar specs \cite{conti1,conti2}.

On the victim radar system, we used all 16 available RX channels and one TX channel. The internal chirp configuration uses $f_s=77GHz$, $b=1GHz$, $t_c=512\mu s$ and a frame size of $N=50$ chirps. This radar system is measuring range and angle-of-arrival using all active receive channels for the purposes of object identification. Signal filtering and object extraction algorithms involve standard OS and CS CFAR implementations, each configured with two guarding cells and $nc=80$ neighbor cells. For the OS-CFAR a scale factor of $sc=0.95$ and an order $k=0.75\cdot nc$ are used \cite{habib2008cfar,sqarelaw}. The CA-CFAR is configured to use a dynamic scaling factor with a false-alarm probability $P_{FA}=0.39$ which yields a similar scaling probability, $sc=0.9528$ (see Section~\ref{sec:cfar}). In addition to the CFAR, radar input signals are also filtered by applying a Hann window.

Our attacking radar system is using the same chirp configuration as the victim (e.g., $f_s=77GHz$, $b=1GHz$, $t_c=512\mu s$) but has no additional knowledge with respect to its internal configuration. The attacker is not synchronized to the victim in any way and is not performing any synchronization procedure prior to the attack. In order to maintain as much control as possible, the attacker is using frame sizes of $N=1$ chirp.

In our experiments, we used digital signal amplification in the range of -5dB to 31dB. As expected, the best results were obtained at the lower end of the range, thus all results should assume a signal strength of -5dB.

\subsection{Baseline attack results}
\label{sec:res_sync}

In this section, we present the fundamental lessons learned from our synchronized adversary. We present this more intuitive scenario first, but highlight the fact that the results for our main contributions are presented in Section \ref{sec:res_non-sync}.

We outlined the theoretical aspects of this attack in previous sections (mainly Section~\ref{sec:sync_def}), and while they seem fairly straight forward, in practice we discovered that achieving perfect synchronization with a victim is a difficult endeavor due to several reasons. First, the main limiting factor is the requirement of specialized hardware which not only supports continuous signal listening in the mmWave bandwidth range (e.g., 76-78GHz), but is also able to behave like a radar unit with respect to generating chirps and computing IFs. To the best of our knowledge, no COTS device of this type currently exists. Mainly due to this first reason, for our real-world experiments, we decided to forgo the complicated procedures outlined in literature \cite{komissarov2021spoofing} and opted instead for an out-of-bounds synchronization via a physical cable between the adversary and the victim radar unit, while maintaining the rest of the interaction between the victim and the attacker in the wireless RF domain. We recognize that this scenario is not suitable for any practical real-world attacks, however it has allowed us to better understand the constraints that affect the signal readings and processing done by radar systems in real-world wireless adversarial settings, and led us to develop our fully wireless non-synchronized attacker model. We continue by detailing the identified challenges for this attacker and our workarounds. 

\subsubsection{Identified attack challenges}

\begin{figure}[tb]
    \centering
    \includegraphics[width=220pt,trim={20 15 15 15},clip]{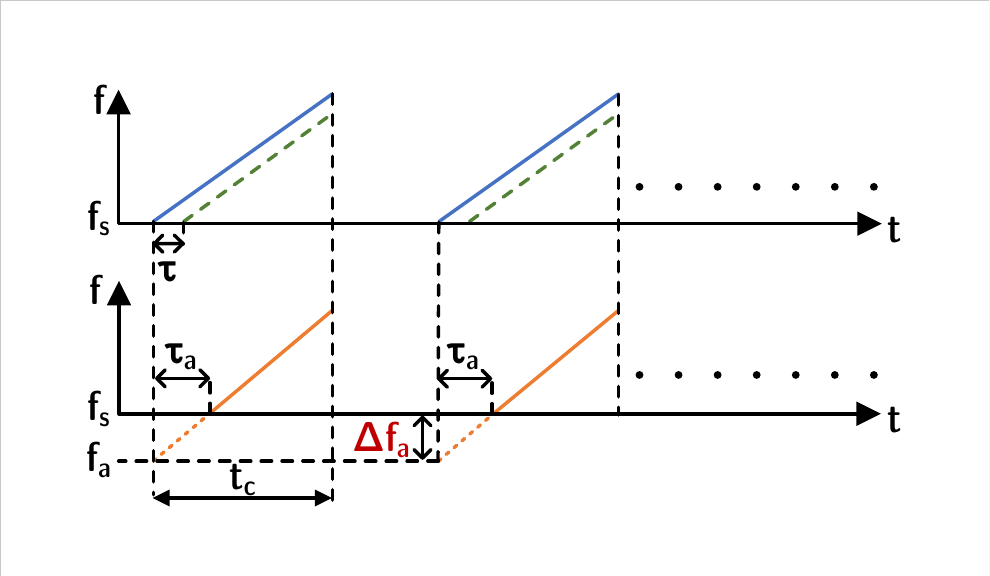}
    \caption{The adversary starts its chirp at the same time as the victim, but from a different, lower frequency $f_a = f_s-\Delta f_a$. The adversary reaches the victim's starting frequency $f_s$ after the intended delay $\tau_a$. The part of the attacker's generated chirp with the frequency below $f_s$ will not be included in the IF computed by the victim, and any signal processing will therefore not be affected by this chirp component.}
    \label{fig:sync_att_freq}
\end{figure}

The main problem we identified is related to the difficulty of maintaining a synchronized state between the victim and the attacker. The limiting factor here is that all electromagnetic waves travel with the same speed, namely the speed of light. This includes radar chirps and reflections, as well as synchronization pulses sent via the synchronization cable. This constraint poses a problem both in our case, and in the case of the classic radio synced attacker, because it forces the adversary to precisely know, and take into account, the distance from the victim in order to compute an accurate \textit{synchronization offset}. More specifically, for a successful attack, the adversary is required to accurately compute the chirp time value $T_c$ as a function dependent on its distance to the victim. This is especially challenging when the attacker and the victim are in relative motion to each other.

Our second immediate observation was that even though we are using identical radar devices for our victim and attacker, the internal clocks they are equipped with are functioning at slightly different frequencies. This manifests as a drift between the two units and requires periodic synchronization pulses, ideally before every chirp. This is important because noise filtering algorithms, e.g., CFAR, will reject signals that are not consistent between multiple chirps and frames. However, in practice, perfect chirp-level synchronization is impossible to achieve due to the tight time constraints related to signal generation and measurement.

Finally, the last issue we identified is related to the adversary's ability to generate chirps with sufficiently small delays, $\tau_a$ (see Section~\ref{sec:sync_def}, Fig.~\ref{fig:sync_att_time}), such that they resemble reflections i.e., they reach the victim during its listening period. Our observations revealed that the timing adjustments available for our COTS radar unit were not precise enough to allow this and enable us to achieve our goals. On our test devices, the victim radar simply rejected/ignored the pulses if the computed attacker delay was $\tau_a \geq 10\mu s$.

\subsubsection{Workarounds}

We designed our workarounds considering an adversary that is only able to use COTS devices. We assume that more efficient methods can be devised to overcome the above-mentioned physical limitations with custom designed hardware.

\begin{figure}[t]
    \centering
    \includegraphics[width=240pt,trim={0 0 0 0},clip]{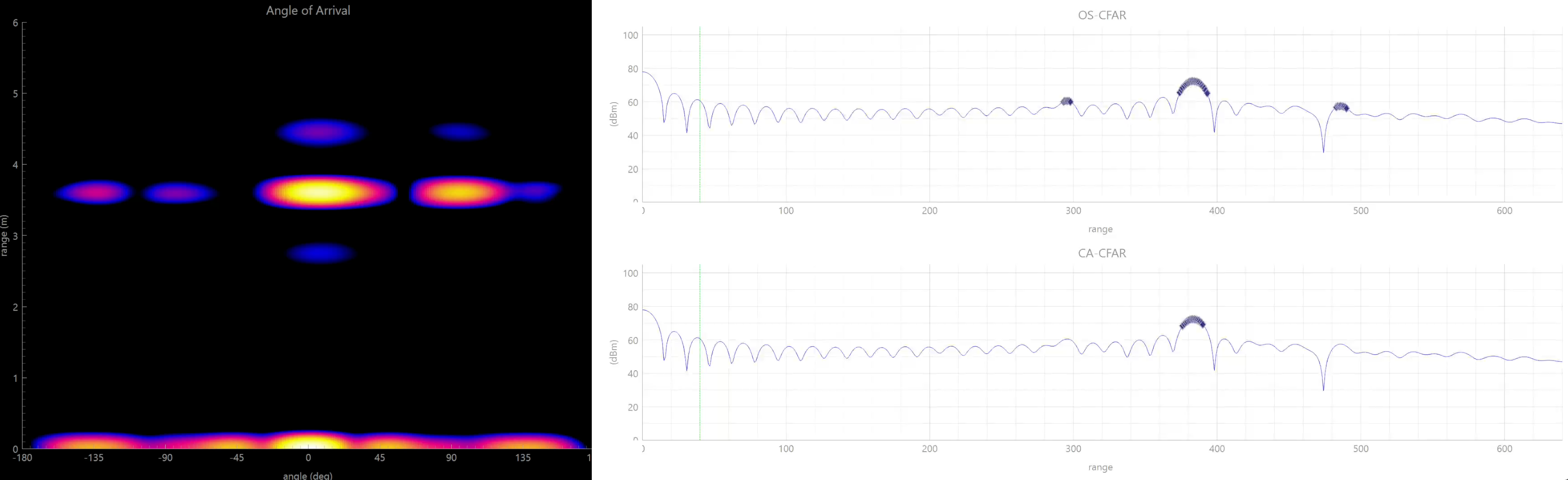}
    \caption{Plot showing angle-of-arrival (left) together with OS-CFAR (top-right) and CA-CFAR (bottom-right) range detection for attacker generated virtual targets. CFAR identified objects are highlighted on the CFAR range plots (i.e., OS-CFAR detected three objects and CA-CFAR one). Some range identified objects are comprised of multiple adjacent objects, as shown on the AoA plot.}
    \label{fig:aoa_cfar}
\end{figure}

We address the synchronization problem by performing frame level synchronization. Even though this solution is not perfect, it allows us to maintain partial synchronization between the two devices with a minimal overhead. With this solution synchronization pulses are sent between frames after which each radar unit observes a fixed time delay (i.e., $200ms$) before starting to send its own chirps. The challenge here is choosing the size of the chirp frame to be sufficiently large such that the sync pulses are minimal, yet small enough such that the drift does not manifest. Through experimentation, we were able to determine that no significant drift appears for frame sizes of $N\le50$ chirps. Additionally, we have observed that if range-Doppler measurements are not performed by the victim radar then \textit{one-chirp frames} can be configured which result in perfect synchronization, however with the caveat of larger delays between chirps and therefore lower real-time tracking precision.

\begin{figure*}[!ht]
    \centering
    \includegraphics[width=\textwidth,trim={10 10 10 10},clip]{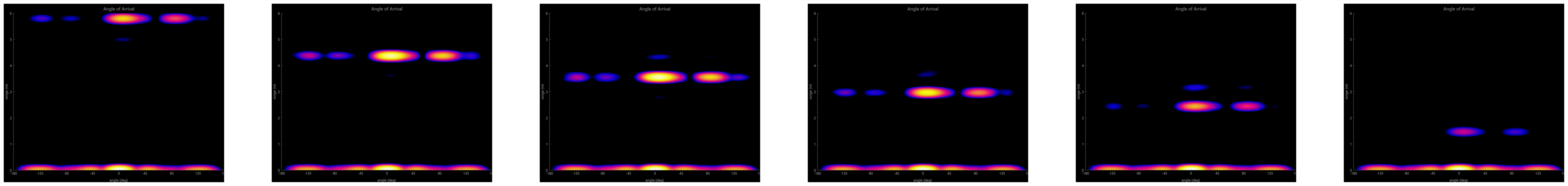}
    \caption{Plots showing a group of virtual objects moving towards the victim radar, i.e., radar front is at the bottom. The two strongest reflections are directly generated by the attacking radar. The secondary reflections are uncontrolled reflections introduced by the environment. The plots shown are sampled from a screen capture recording of the attack.}
    \label{fig:moving_aoa}
\end{figure*}

Given that we are able to synchronize the two boards, we decided to leverage this when addressing the issue related to the precision of the adversary's generated signal delay. Our solution is outlined in Fig.~\ref{fig:sync_att_freq} and works as follows. For a chosen fixed delay $\tau_a$, and a chirp slope $S$, the adversary computes a new starting frequency $f_a$ lower than the victim's starting frequency $f_s$ such that $f_s=f_a+S\cdot\tau_a$. The synchronized attacker then proceeds to send its chirps starting from $f_a$ at the same time as the victim. This gives the attacker a frequency-based control of the delay value $\tau_a$ without adding any noise due to the fact that frequencies outside the range $[f_s,f_s+S\cdot t_c)$ are automatically rejected by the victim radar. We are also able to leverage this finding for our non-synchronized adversary in order to better control the movement of virtual objects. 

\subsection{Real-world attack results}
\label{sec:res_non-sync}

Prior research \cite{komissarov2021spoofing} suggests that synchronization is required to generate virtual objects. In this section we present our results which show that this, in fact, is not the case. Basic knowledge of the chirp parameters ($f_s$, $b$ and $t_c$) is enough to generate targets in a victim radar without requiring any additional synchronization feedback. In order to successfully deploy our non-synchronized attack, we leverage the theoretical concepts detailed in Section~\ref{sec:theo_nonsync} as well as the insights from our baseline synchronized attacker. In our evaluation, we found that the attack procedure presented in the following is deterministic on the devices tested.

We begin with an example of our attack which generates two virtual objects as shown in Fig.~\ref{fig:aoa_cfar}. The these results demonstrate the capabilities of a non-synchronized adversary which uses two transmit antennas to generate two adjacent objects placed in the same range bin. This adversary is also raising the ambient noise levels to remove most of the physical objects present in the environment. We further explain these findings below.

\textit{Constructive noise generation.} Using the example above, we analyze the difficulty of generating useful noise, for the purposes of hindering measurements on the victim device, and obfuscating real physical objects. The first goal is to determine the minimum and maximum frequency, relative to the victim's $f_s$ and $b$, that the attacker can use to influence the victim's IF. Our approach is different to standard noise attack methods, which tend to blanket the victim's full radio spectrum. The problem with full spectrum blanking is that reflections are no longer detected at all due to high environmental noise levels \cite{ewf,electronics9040573}. So, through experimentation, we determine that variations to the chirps starting frequency, $f_s$, with $\pm10^5Hz$ outside the configured chirp frequency range (i.e $77\cdot10^{9}\pm10^{5}Hz$) do not interfere at all with the victim radar and are, in fact, filtered out by the hardware signal filters and internal signal processing algorithms, rendering them useless for the purposes of a constructive noise attack. The cause for this behavior is that a victim's listening period ends with the end of its TX chirp and, given longer time intervals are required to reach the higher frequencies, they end up not being sampled. As such, we bound the search space for the injected chirp's starting frequency to the above interval and determine hard bounds for the $\Delta f_a$ parameter (see Fig.~\ref{fig:sync_att_freq}). In turn, we observe an overall increase in the victim's perceived noise levels, and significant alterations of the victim's IF, produced by the injected chirps that were restricted to the appropriate interval (e.g. $\pm 0.1MHz$), shown in Fig~\ref{noise_fft}. This, however, does not completely remove strong reflective objects (e.g. corner reflectors) from the victim's view. It is enough, though, to bias the CFAR algorithm baseline such that when even stronger reflections (i.e. virtual objects) are placed in the view they end up obfuscating the remaining physical objects. We detail our virtual object generation next.  

\textit{Virtual target generation.} For our virtual object generation, the goal is to find suitable values for the attacker chirp period $T_a$. We observed a successful attack strategy that starts with the attacker computing a chirp period that is a denominator to the victim's $T_c$ period. Then the attacker can freely choose any $T_a$ value strictly smaller than the previously computed value. This configuration allows the most flexibility for the attacker, but requires victim chirps to be loosely spread out within the frame (or an attacker with more powerfully hardware). Using this strategy, we were able to generate virtual objects that are positioned both in front and behind the physical attacking radar device. We further experiment with the limit case where the attacker chirp period is equal to the victim, $T_a=T_c$. We observe that this type of attack is also possible. The chirps tend to converge but only after long periods of time, which, as expected, are closely related to the initial alignment, i.e., the time difference between the victim's chirp and the attacker's chirp at the moment the attacker starts transmitting for the first time. We speculate that this convergence is possible due to the internal clock drift between the two units.

\textit{Multiple object generation.} In our last experiment, we look at the impact of using several TX antennas on the attacker radar to generate multiple virtual objects. We were able to confirm our hypothesis outlined in Section~\ref{sec:multi_obj}.

\begin{figure}[t]
    \centering
    \includegraphics[width=250pt,trim={0 0 0 0},clip]{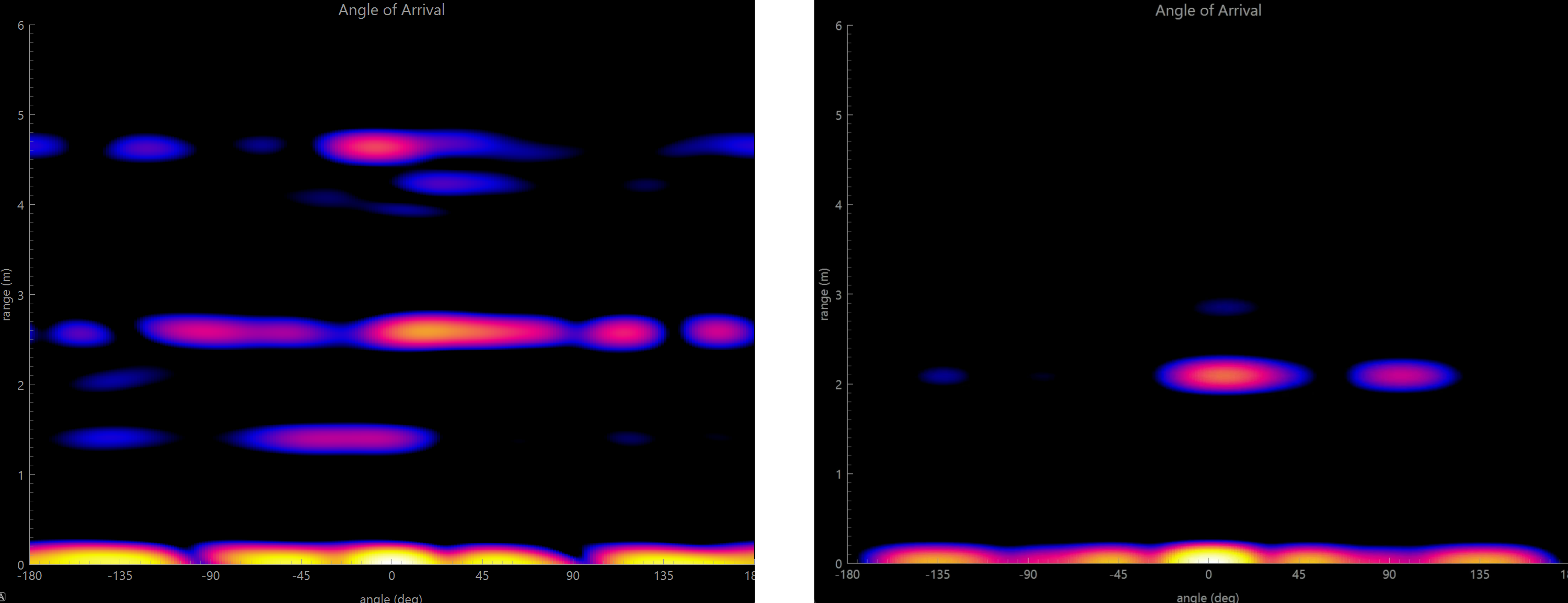}
    \caption{Constructive object removal. Left: AoA view of the victim without any interference from the attacker. Strongest reflections from top-to-bottom are: T1, Corner reflector and T2 (CFAR view is shown in Fig.~\ref{fig:cfar_graph}). Right: AoA view of the victim with constructive object removal and target generation. Strongest reflections represent virtual objects.}
    \label{fig:noise_cfar}
\end{figure}

\textit{Moving virtual objects.} Our results are presented in Fig.~\ref{fig:moving_aoa}. We are able to generate several moving objects positioned at different angles with respect to the victim radar using the antennas on the attacking radar device. The virtual objects shown are moving towards the victim radar with a velocity of $\approx 0.016 m/s$ ($\approx0.5km/h$). As can also be seen, the attacker injected signal is also raising the overall noise levels observed by the victim, which removes the physical objects from the scene. Fig.~\ref{fig:noise_cfar} shows the victim's view before the attack (left), when a multitude of objects and reflections are detected, and during the attack (right), when only the virtual generated objects are identified.

\section{Discussion and Related Work}
Spoofing attacks against radar systems have been known to exist for quite some time \cite{katsilieris2013detection}, however prior to this work no spoofing attack has been demonstrated against radar systems deployed in a real-world, unrestricted environment. We give a summary of recent work in the following.

In \cite{komissarov2021spoofing} Kommissarov at al. propose an attack against automotive FMCW mmWave radars using an emulated environment comprised of two bladeRF SDRs connected by 15m-long RF cables. Their attack consists of two phases: a synchronization phase based on signal time-of-arrival, and a signal injection phase. Using simulations, the authors show that by spoofing the distance and velocity, they can trick a victim into detecting a phantom stop or acceleration. They evaluate several known, \textit{but not in-use}, countermeasures such as sensor fusion with LiDAR/camera, phase randomization and frequency randomization, against their attacks.

In \cite{nashimoto2021low, ashes_19} authors present a low-cost distance-spoofing attack in using a
wired connection between the victim and the attacker. The victim is a 24GHz mmWave FMCW radar using a VCO ADF5901 module, a triangular wave generator ADF4159, and a down conversion mixer ADI ADF5904. The attacking device is an Arduino Due module. A half-chirp modulation scheme is used to overcome crystal oscillators timing constraints, while a two-step delay insertion scheme precisely controls relative delay difference between the participants. Counter-measures are also proposed in the form of random-chirp modulation and its security level is evaluated under the proposed attack.

In \cite{tmtt_21} authors develop a hardware based spoofing device that uses a single-sideband mixer to introduce a frequency shift to the incoming victim RF signal before retransmitting it. Using a prototype in a non-standard radar frequency (e.g., 5.8GHz) authors demonstrate that the modulated RF signal creates an illusion of a real target in the victim. They also propose a hybrid-chirp approach to distinguish real targets from a spoofed one.

In \cite{vtc_18} authors propose a spatio-temporal challenge-response (STCR) method to detect and mitigate sensor spoofing attacks. SCTR works by dividing the radar bandwidth into narrower beams and randomly selecting one for each transition. Reflections that cannot be paired with a transmitted beam are considered attack signals and are rejected by the radar unit. The solution has only been implemented in software using an adaptive cruise control package in Matlab.

In \cite{moon2020bluefmcw} authors introduce a FMCW radar scheme called BlueFMCW that mitigates both interference and spoofing signals by randomly hopping frequency. They are able to maintain high radar resolution and protect against phase discontinuity using their proposed phase alignment algorithm. They support their findings using simulations.

\subsection{Radar object spoofing mitigations}
Based on the attacks we have demonstrated, and the partial results already available in the state of the art, we consider that it is important that mitigations against spoofing attacks should be added to radar systems. Solutions based on randomization such as the ones reviewed in \cite{komissarov2021spoofing, nashimoto2021low} could be efficient against our attacks. In the following, we briefly list the most relevant of these and reason about their efficiency.

\textit{Frequency and Phase Randomization.}
As we have shown previously, in order to successfully execute our attacks and generate artificial objects an adversary needs to have some prior knowledge about the victim's operating parameters, e.g., starting frequency, bandwidth and chirp duration, thus attacks can be prevented by hiding these parameters. This can be achieved by randomizing the parameters of chirps. Unfortunately, bandwidth and chirp duration directly influence the measurement by affecting the resolution and the freshness, respectively. Thus, the only viable candidate remains the starting frequency. In \cite{moon2020bluefmcw, nashimoto2021low, ashes_19} a couple of solutions are proposed to implement frequency randomization, results about their effectiveness is only provided through numeric simulations. Real-world implementations of these countermeasures would require additional hardware, however, if proven to work they would make our attacks impossible to deploy.

Another chirp related parameter that can be randomized is the phase of the signal. This has been mentioned in \cite{komissarov2021spoofing} as an effective countermeasure to their simulated velocity spoofing attack, which requires the victim radar to receive a signal with the same phase as the one transmitted. The authors note that complex circuity would be required for implementing it, and mention that the process could be very error-prone. Based on our experiments, we do not expect this countermeasure to be particularly efficient against our attacks. As we have shown in Section~\ref{sec:multi_obj} phase variations directly affects both velocity and AoA measurements in ways that enhance the attack. As such, in order to be effective as a countermeasure, phase randomization needs to be paired with additional signal filtering, further increasing the complexity of the system.

\textit{RSSI Measurements.} Our non-synchronized attacks unavoidably affect the signal-noise in the environment. A Received Signal Strength Indicator (RSSI) based countermeasure could prove effective in detecting our attacks if the victim is able to differentiate between the normal signal power of its reflections and the signal power of the injected signal, which should be elevated in the area of the attack. Another, more advanced technique, is proposed in \cite{komissarov2021spoofing}, where the victim attempts to detect patterns of $n$ weak chirps out of the total $N$ chirps in the frame. This technique however might fail as they uniformly raise the noise level in the environment, effectively cancelling out all the victim's reflections.

\textit{LiDAR/Camera sensor integration.} Multi-sensor integration is another solution that has been proposed in literature as a potential countermeasure to radar object-spoofing attacks \cite{yang2021secure}. This works by measuring the same physical variables in the environment using different sensors and verifying the measurements between these. This method however is unlikely to be effective in practice due to safety requirements, which currently favor combining sensors using logical OR functions rather than AND functions. These requirements ensure that if a dangerous situation is encountered, and is then identified even by just one of the sensors, then the safe action is executed and not ignored.

\section{Conclusions}
FMCW radar devices are an integral part of the automotive safety-critical ADAS, and often the only reliable sensor. Due to their hardware complexity they have, until now, only been evaluated in constrained or simulated environments. In this work, we reviewed some of the low-level functional details of these devices and identified several methods in which they can be exploited by an adversary. We have demonstrated several object spoofing attacks against automotive-grade radars deployed in truly wireless environments. We have shown that synchronization with the victim is not required for a successful attack and can, in fact, help the adversary. On our radar devices, we have also shown that object removal through noise is a trivial endeavor if the attacker mimics the chirps of the victim and does not exceed the chirp's bandwidth by more than $0.1MHz$. Given the relatively low-bar required to deploy these attacks, and the importance of this safety critical component, we believe that radar manufacturers and automotive industry stake-holders should consider integrating efficient object spoofing countermeasures, such as randomized frequency hopping, as a matter of urgency.

\bibliographystyle{ieeetr}
\bibliography{IEEEabrv,bibliography}

\end{document}